\documentclass[reprint, aps, pra, amsmath, amssymb,superscriptaddress]{revtex4-2}
\usepackage{graphicx}
\usepackage{placeins}
\usepackage[colorlinks=true, linkcolor=blue, citecolor=blue, urlcolor=blue]{hyperref}
\allowdisplaybreaks[4]
\begin{document}

\title{Phase-Selected Efficient Single-Photon Frequency Conversion via Local Fano Resonance in a Two-Giant-Atom Waveguide-QED System}

\author{Qing-Ao Xiang}
\affiliation{Key Laboratory of Opto-electronic Control and Detection Technology of University of Hunan Province, and College of Physics and Electronic Engineering, Hengyang Normal University, Hengyang 421002, China}

\author{Yan Liu}
\affiliation{Key Laboratory of Opto-electronic Control and Detection Technology of University of Hunan Province, and College of Physics and Electronic Engineering, Hengyang Normal University, Hengyang 421002, China}

\author{Xin-Yuan Yang}
\affiliation{Key Laboratory of Opto-electronic Control and Detection Technology of University of Hunan Province, and College of Physics and Electronic Engineering, Hengyang Normal University, Hengyang 421002, China}

\author{Ya-Ju Song}
\email{yjsong@hynu.edu.cn}
\affiliation{Key Laboratory of Opto-electronic Control and Detection Technology of University of Hunan Province, and College of Physics and Electronic Engineering, Hengyang Normal University, Hengyang 421002, China}
\affiliation{Key Laboratory of Low Dimensional Quantum Structures and Quantum Control of Ministry of Education，Hunan Normal University, Changsha 410081, China}
\date{\today}

\begin{abstract}
Efficient single-photon frequency conversion is investigated in a two-giant-atom waveguide-QED system, where a two-level giant atom and a $\Lambda$-type three-level giant atom couple to a common one-dimensional waveguide. While the $\Lambda$-type atom provides the inelastic channel, the two-level atom induces secondary coherent coupling, creating multi-path interference for the converted photon. Using the real-space approach and within the Markovian approximation, we derive analytical four-channel scattering amplitudes and reveal that the inelastic transmission spectrum, governed by three complex resonance poles, exhibits a multi-peak interference pattern. By introducing a local single-pole approximation, we reduce this complex spectrum to a local Fano lineshape, decomposing it into a coherent superposition of a local background term and a single-pole resonant term. The interplay between these two terms—controlled by the photon propagation phase between the giant atoms' coupling points—determines the conversion efficiency, with the background suppression condition leading to a Lorentzian reduction. Based on the single-pole resonance weight, we formulate a phase-selection criterion for highly efficient conversion. Compared with both the small-atom and single $\Lambda$-type giant-atom models, the two-giant-atom scheme achieves substantially enhanced inelastic transmission over a broader frequency-conversion range. This work reveals how phase-controlled local Fano resonance enables high-efficiency frequency conversion, establishing a general paradigm for engineering resonant light-matter interactions in structured quantum systems.
\end{abstract}
\maketitle

\section{Introduction}

Waveguide quantum electrodynamics (waveguide QED) studies the coherent interaction between propagating photons and quantum emitters in one-dimensional waveguides~\cite{Kimble2008QuantumInternet,ShenFan2005SinglePhoton,Shen2007}. Unlike cavity QED with discrete modes, waveguide QED supports a continuum of traveling-wave photons. Strong transverse confinement enhances the emitter-field coupling, and the continuum nature enables versatile control over interference effects in single-photon scattering, quantum state transfer, photon routing, and frequency conversion~\cite{Roy2017Colloquium,Sheremet2023WaveguideQED,Kannan2023,Gonzalez-Tudela2024,Liao2016Review}. These features establish waveguide QED as a crucial platform for investigating light-matter interactions at the single-photon level and for advancing quantum information processing technologies. Realizations include quantum dots coupled to plasmonic nanowires~\cite{Akimov2007}, photonic crystal waveguides~\cite{Arcari2014}, natural atoms in optical fibers~\cite{Sayrin2015}, and superconducting qubits coupled to transmission lines~\cite{Yoshihara2016}.

In conventional waveguide-QED systems, quantum emitters are typically treated as point-like small atoms, with dimensions much smaller than the photon wavelength, and thus couple to the waveguide at a single spatial position. Recently, artificial quantum systems—such as superconducting circuits and surface-acoustic-wave devices—have enabled giant-atom structures where atoms are coupled to propagating modes at multiple spatially separated positions. Furthermore, the level structures of these multilevel artificial atoms can be flexibly engineered~\cite{Gu2017MicrowavePhotonics,Blais2021CircuitQED,Gustafsson2014SAW,Kockum2021GiantAtomsReview,Vadiraj2021LevelStructure,Andersson2019Nonexponential,Kockum2014GiantAtom,Kannan2020GiantAtoms}. Crucially, photon propagation between these coupling points accumulates phases, rendering emission, absorption, and scattering jointly determined by local coupling strengths, propagation phases, and multipath interference. This multipoint coupling mechanism provides new degrees of freedom for designing quantum devices with tunable spectral structures and highly selective scattering properties~\cite{Kockum2014GiantAtom,Kannan2020GiantAtoms,CaiJia2021GiantScattering,Feng2021TwoGiantAtoms,Kockum2018DecoherenceFree,Zhao2020MultipleCouplingPoints,Zou2022GiantLambdaSQUID,Peng2023ChainGiantAtoms,Chen2022NonreciprocalGiantAtoms,Andersson2019Nonexponential,Yin2022GiantMolecule}.

Single-photon frequency conversion serves as a crucial interface for connecting quantum nodes operating at different frequencies~\cite{Kimble2008QuantumInternet,Raymer2012SinglePhotons,Bradford2012SagnacConversion,Maring2018MemoryConversion}.
Within the framework of waveguide QED, this process is realized through inelastic scattering from multilevel quantum emitters: an incident photon exchanges energy with internal atomic transitions and is re-emitted into the waveguide at a different frequency~\cite{Tsoi2009LambdaScattering,Witthaut2010ThreeLevelEmitter,Bradford2012FrequencyConversion,Wang2014FrequencyConverter,Du2021GiantLambdaConversion,Du2021ChiralConversion,Huang2026GiantThreeLevelConversion,Li2026TShapeConversion}. How to utilize quantum interference to enhance the target inelastic transmission channel while suppressing other elastic or backward scattering channels is a key issue for achieving efficient frequency conversion.

In this work, we investigate a waveguide-QED structure comprising a two-level giant atom and a $\Lambda$-type three-level giant atom, both coupled to a common waveguide via two spatially separated points. The $\Lambda$-type atom provides the frequency-conversion channel, while the two-level atom induces secondary coherent coupling that creates multipath interference for the converted photon. We demonstrate that propagation-phase control over this interference achieves high-efficiency single-photon frequency conversion. Specifically, we focus on the forward inelastic transmission, as it directly corresponds to the targeted output for quantum routing and provides the most transparent physical picture for the enhancement mechanism; notably, the inelastic reflection is correspondingly suppressed under optimal phase conditions.
Using the real-space approach~\cite{ShenFan2005SinglePhoton,Zhou2008ControllableScattering} and within the Markovian approximation~\cite{Guo2017GiantAtom,Sinha2020NonMarkovian}, we derive analytic expressions for the four-channel scattering amplitudes (elastic and inelastic transmission and reflection). Under the equal-spacing and symmetric-coupling conditions, the inelastic transmission spectrum exhibits a multi-peak interference structure induced by three complex resonance poles. Around the target scattering window, we construct a local single-pole approximation and decompose the inelastic transmission amplitude into a coherent superposition of a local background term and a single-pole resonant term, yielding a local Fano lineshape~\cite{Fano1961Configuration,Miroshnichenko2010Fano,Limonov2017Fano,Feng2021TwoGiantAtoms,Yin2022GiantMolecule}. 
We then use the single-pole resonance weight as a phase-selection criterion. 
The enhancement of the inelastic transmission probability within the target scattering window is verified using the exact four-channel scattering probabilities. 
The frequency-conversion advantage brought by secondary coherent coupling and internal propagation-phase control is assessed through comparison with the small-atom and single $\Lambda$-type giant-atom models. Furthermore, by introducing excited-state loss, we analyze the robustness of the system against loss.

The paper is organized as follows. Sec.~\ref{sec:model} introduces the physical model and solves the four-channel scattering amplitudes using the real-space approach. Sec.~\ref{sec:SPA} analyzes the local single-pole approximation and Fano lineshape in the inelastic transmission spectrum. Sec.~\ref{sec:results} combines the model comparison, frequency-conversion interval dependence, and robustness analysis under excited-state loss. Sec.~\ref{sec:conclusion} presents the summary and discussion.

\FloatBarrier
\section{Two-Giant-Atom Model and Analytical Four-Channel Scattering}
\label{sec:model}
\begin{figure}[htbp]
    \centering
    \includegraphics[width=0.8\columnwidth]{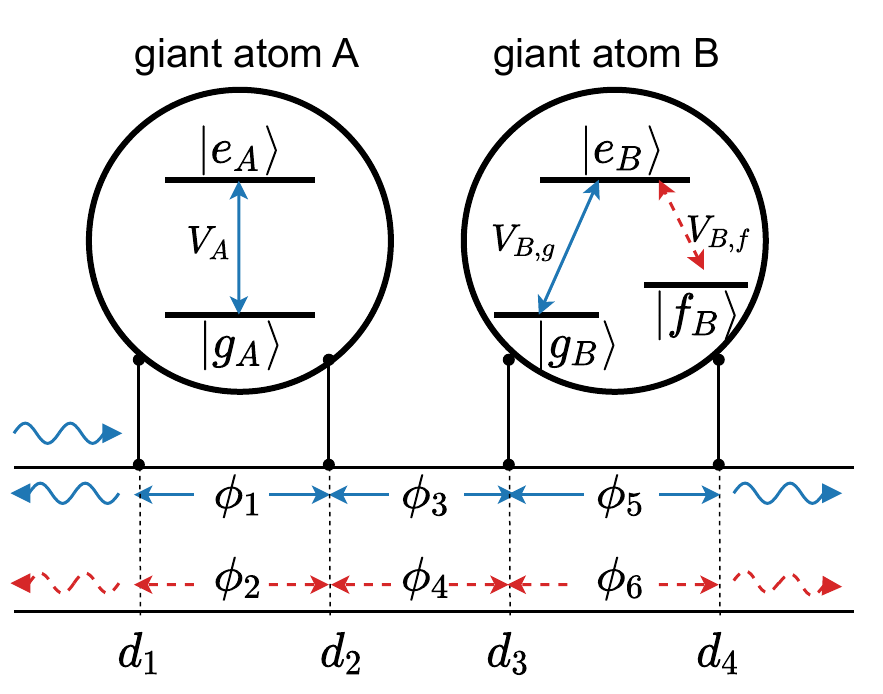}
    \caption{Schematic of the waveguide-QED system. A two-level giant atom $A$ and a $\Lambda$-type three-level giant atom $B$ couple to a common waveguide at points $d_{1,2}$ and $d_{3,4}$, respectively. The blue solid and red dashed lines denote the g- and f-channel photons, with propagation phases $\phi_{2n-1}=k_{g}(d_{n+1}-d_{n})$ and $\phi_{2n}=k_{f}(d_{n+1}-d_{n})$, $n=1,2,3$.}
    \label{fig:TGA-Schematic}
\end{figure}

As shown in Fig.~\ref{fig:TGA-Schematic}, we consider a one-dimensional waveguide-QED system composed of a two-level giant atom $A$ and a $\Lambda$-type three-level giant atom $B$. The two atoms are coupled to the waveguide at spatially separated points, with coupling coordinates specified by $d_{1}$ and $d_{2}$ for atom $A$, and $d_{3}$ and $d_{4}$ for atom $B$~\cite{Kockum2014GiantAtom,Kannan2020GiantAtoms,Vadiraj2021LevelStructure,Feng2021TwoGiantAtoms,Zhao2020MultipleCouplingPoints,Du2021GiantLambdaConversion,CaiJia2021GiantScattering,Zou2022GiantLambdaSQUID,Peng2023ChainGiantAtoms,Kockum2018DecoherenceFree}. Throughout this work, we set $\hbar=1$.
The total Hamiltonian of the system is given by ($\hbar=1$)~\cite{ShenFan2005SinglePhoton,Zhou2008ControllableScattering}
\begin{equation}
\hat{H}=\hat{H}_w+\hat{H}_a+\hat{H}_{\mathrm{int}},
\end{equation}
where $\hat{H}_w$ is the free Hamiltonian of the waveguide, $\hat{H}_a$ describes the two giant atoms, and $\hat{H}_\mathrm{int}$ represents their interaction. The free waveguide Hamiltonian $\hat{H}_w$ can be expressed as
\begin{align}
    \hat{H}_w &=\int dx\bigg[\hat{a}_R^\dagger(x)\left(\omega_0-iv_g\frac{\partial}{\partial x} \right)\hat{a}_R(x)\nonumber\\
    &\quad+\hat{a}_L^\dagger(x) \left( \omega_0+iv_g\frac{\partial}{\partial x}\right)\hat{a}_L(x) \bigg],
\end{align}
where the integration is taken from $-\infty$  to $+\infty$ over the waveguide coordinate $x$, $\hat a^\dagger_{R,L}(x)$ ($\hat a_{R,L}(x)$) are the creation (annihilation) operators for right- and left-propagating photons, respectively, $v_g$ is the waveguide group velocity, and $\omega_0$ is the reference frequency for the linearized waveguide dispersion.
The free Hamiltonian of the two giant atoms is
\begin{equation}
    \hat{H}_a =\omega_e \left( \vert e_A\rangle\langle e_A\vert +\vert e_B\rangle\langle e_B\vert \right) + \omega_f \vert f_B\rangle\langle f_B\vert,
\end{equation}
where we take the energies of the ground states $|g_A\rangle$ and $|g_B\rangle$ as the zero reference, setting $\omega_g=0$. The excited-state energies of both giant atoms are $\omega_e$, and the energy of the metastable state $|f_B\rangle$ of atom $B$ is $\omega_f$, which defines the frequency-conversion interval. The interaction Hamiltonian is
\begin{align}
\hat{H}_{\mathrm{int}}
&=\int dx\Bigl\{\bigl[\hat{a}^{\dagger}_{R}(x)+\hat{a}^{\dagger}_{L}(x)\bigr]\nonumber\\
&\quad\times\bigl[S_A(x)\sigma_A^-+\sum_{\alpha=g,f}S_{B,\alpha}(x)\sigma_{B,\alpha}^-
\bigr]
+H.c.\Bigr\}.
\label{eq:H_int}
\end{align}
Here, the values $\alpha=g$ and $\alpha=f$ correspond to the states $\vert g_B\rangle$ and $\vert f_B\rangle$, respectively. The operator $\sigma_A^-=\vert g_A\rangle\langle e_A\vert$ represents the lowering operator for giant atom $A$, while $\sigma_{B,\alpha}^-=\vert\alpha_B\rangle\langle e_B\vert$ denotes the lowering operator associated with the $\vert e_B\rangle\rightarrow\vert\alpha_B\rangle$ transition in giant atom $B$. The spatial coupling distribution between the giant atoms and the waveguide are characterized by $S_A(x)=V_A[\delta(x-d_1)+\delta(x-d_2)]$ and $S_{B,\alpha}(x)=V_{B,\alpha}[\delta(x-d_3)+\delta(x-d_4)]$ for atoms $A$ and $B$, respectively. $V_{A}$ and $V_{B,\alpha}$ denote the respective coupling strengths. 

In the single-excitation subspace, the stationary-state wave function of the system can be written as~\cite{ShenFan2005SinglePhoton,Tsoi2009LambdaScattering,Witthaut2010ThreeLevelEmitter}
\begin{align}
    |\psi\rangle
    &=\int dx \sum_{\alpha=g,f}
[R_\alpha(x)\hat{a}_R^\dagger(x)+L_\alpha(x)\hat{a}_L^\dagger(x)]\vert0,g_A,\alpha_B\rangle
   \nonumber \\
    &\quad+\sum_{\alpha=g,f}u_{A,\alpha}\vert0,e_A,\alpha_B\rangle+u_B\vert0,g_A,e_B\rangle.
\end{align}
Here, $R_\alpha(x)$ and $L_\alpha(x)$ are the spatial amplitudes of right- and left-propagating photons, respectively, and $u_{A,\alpha}$ and $u_B$ denote the excitation amplitudes of the atomic states.
Substituting the wave function into the stationary Schrödinger equation $\hat{H}\vert\psi\rangle=E\vert\psi\rangle$ yields
\begin{equation}
\begin{aligned}
&E R_\alpha(x) = D_\alpha^{-} R_\alpha(x) +S_A(x)u_{A,\alpha}+S_{B,\alpha}(x)u_B, \\
&E L_\alpha(x) = D_\alpha^{+}L_\alpha(x) +S_A(x)u_{A,\alpha}+S_{B,\alpha}(x)u_B, \\
&E u_{A,\alpha} = (\omega_e+\omega_\alpha)u_{A,\alpha} +\int dx S_A^*(x)[R_\alpha(x)+L_\alpha(x)], \\
&E u_B = \omega_e u_B +\int dx \sum_{\alpha=g,f}S_{B,\alpha}^*(x)[R_\alpha(x)+L_\alpha(x)],
\end{aligned}
\end{equation}
where $E$ denotes the total energy of the single-excitation scattering state, and $D_\alpha^{\pm}=(\omega_0+\omega_\alpha\pm iv_g\partial_x)$ with $\omega_{\alpha}$ defined as zero for the ground state ($\alpha=g$) and $\omega_{f}$ for the metastable state ($\alpha=f$). For a single photon incident from the left, the right- and left-propagating probability amplitudes in the waveguide are given as~\cite{ShenFan2005SinglePhoton,Feng2021TwoGiantAtoms}
\begin{equation}
\begin{aligned}
 R_\alpha(x)&=  e^{ik_\alpha x}\bigg\{\delta_{\alpha g}\Theta(d_1-x)+t_\alpha\Theta(x-d_4)\\
&\quad+\sum_{i=1}^{3} M_{i\alpha}\Big[\Theta(x-d_i)-\Theta(x-d_{i+1})\Big]\bigg\},\\
L_\alpha(x)&=e^{-ik_\alpha x}\bigg\{r_\alpha\Theta(d_1-x)\\
&\quad+\sum_{i=1}^{3}N_{i\alpha}\Big[\Theta(x-d_i)-\Theta(x-d_{i+1})\Big]\bigg\}.
\end{aligned}
\end{equation}
Here, $\delta_{\alpha g}$ is the Kronecker delta function, and $k_\alpha$ is the propagation constant satisfying $k_f=k_g-\omega_f/v_g$~\cite{Tsoi2009LambdaScattering,Du2021GiantLambdaConversion}. The quantities $t_\alpha$ and $r_\alpha$ denote the transmission and reflection amplitudes, respectively, with $\alpha=g$ corresponding to the elastic channel and $\alpha=f$ to the inelastic channel. The quantities $M_{i\alpha}$ and $N_{i\alpha}$ represent the right- and left-propagating amplitudes in each interval between adjacent coupling points.
With the detuning defined as $\Delta=E-\omega_e$, the four scattering amplitudes obtained via algebraic elimination are
\begin{equation}
\begin{aligned}
t_g=\frac{\mathcal{N}_{t_g}}{\mathcal{D}},\quad
r_g=\frac{\mathcal{N}_{r_g}}{\mathcal{D}},\quad
t_f=\frac{\mathcal{N}_{t_f}}{\mathcal{D}},\quad
r_f=\frac{\mathcal{N}_{r_f}}{\mathcal{D}}.\label{eq:scattering_amplitudes}
\end{aligned}
\end{equation}
In the absence of loss, the scattering amplitudes satisfy probability conservation: $|t_g|^2+|r_g|^2+|t_f|^2+|r_f|^2=1$~\cite{Bradford2012FrequencyConversion}. The common denominator in Eq.~\eqref{eq:scattering_amplitudes} takes the form
\begin{equation}
\mathcal{D}=\mathcal{E}_g\mathcal{E}_f\mathcal{E}_B+\mathcal{G}_g\mathcal{E}_f+\mathcal{G}_f\mathcal{E}_g,
\label{eq:D_common}
\end{equation}
with the respective numerators
\begin{widetext}
\begin{equation}
\begin{aligned}
\mathcal{N}_{t_g}
&=(\Delta-2\Gamma_A\sin\phi_1)\bigg\{\mathcal{E}_f\Big[\Delta-2\Gamma_{B,g}\sin\phi_5+2i\Gamma_{B,f}(1+e^{i\phi_6})\Big]+\mathcal{G}_f\bigg\}.\\
\mathcal{N}_{r_g}&=-i\bigg\{\Gamma_A(1+e^{i\phi_1})^2(\mathcal{E}_f\mathcal{E}_B+\mathcal{G}_f)+e^{2i(\phi_1+\phi_3)}(1+e^{i\phi_5})^2\Gamma_{B,g}\mathcal{E}_f\Big[\Delta-2i\Gamma_A(1+e^{-i\phi_1})\Big]\bigg\}.\\
\mathcal{N}_{t_f}&=-e^{i(\phi_1-\phi_2+\phi_3-\phi_4-\phi_6)}(1+e^{i\phi_5})(1+e^{i\phi_6})\sqrt{\Gamma_{B,g}\Gamma_{B,f}}(\Delta-2\Gamma_A\sin\phi_1)\\
&\quad\times\bigg[e^{i(2\phi_4+\phi_6)}(1+e^{i\phi_2})^2\Gamma_A-2\Gamma_A(1+e^{i\phi_2})+i(\Delta-\omega_f)\bigg].\\
\mathcal{N}_{r_f}&=-e^{i(\phi_1+\phi_2+\phi_3+\phi_4)}(1+e^{i\phi_5})(1+e^{i\phi_6})\sqrt{\Gamma_{B,g}\Gamma_{B,f}}(\Delta-2\Gamma_A\sin\phi_1)\Big[i(\Delta-\omega_f)-2i\Gamma_A\sin\phi_2\Big].
\label{eq:numerators}
\end{aligned}
\end{equation}
\end{widetext}
The auxiliary quantities in the above expressions are
\begin{equation}
\begin{aligned}
\mathcal{E}_g&= \Delta+2i\Gamma_A(1+e^{i\phi_1}),\\
\mathcal{E}_f&=\Delta-\omega_f+2i\Gamma_A(1+e^{i\phi_2}),\label{eq:auxiliary_definitions}\\
\mathcal{E}_B&=\Delta+2i\Gamma_{B,g}(1+e^{i\phi_5})+2i\Gamma_{B,f}(1+e^{i\phi_6}),\\
\mathcal{G}_g&=e^{2i\phi_3}(1+e^{i\phi_1})^2(1+e^{i\phi_5})^2\Gamma_A\Gamma_{B,g},\\
\mathcal{G}_f&=e^{2i\phi_4}(1+e^{i\phi_2})^2(1+e^{i\phi_6})^2\Gamma_A\Gamma_{B,f}.
\end{aligned}
\end{equation}
Here, $\mathcal{E}_g$, $\mathcal{E}_f$, and $\mathcal{E}_B$ are complex detunings governed by the two-point coupling phases. Specifically, $\mathcal{E}_g$ and $\mathcal{E}_f$ characterize the coherent coupling of the $g$- and $f$-channel photons to giant atom $A$ at points $d_{1}$ and $d_{2}$, respectively, while $\mathcal{E}_B$ characterizes their coupling to giant atom $B$ at points $d_{3}$ and $d_{4}$. The $\mathcal{G}_g$ and $\mathcal{G}_f$ represent the coherent coupling between the two giant atoms mediated by the $g$- and $f$-channel waveguide modes, respectively~\cite{vanLoo2013Photon,Lalumiere2013InputOutput,Zheng2013Waveguide}. The decay rates for the respective transition paths are defined as
\begin{equation}
\Gamma_A=\frac{|V_A|^2}{v_g},
\quad
\Gamma_{B,g}=\frac{|V_{B,g}|^2}{v_g},
\quad
\Gamma_{B,f}=\frac{|V_{B,f}|^2}{v_g},
\end{equation}
where $\Gamma_A$ is associated with the $|g_A\rangle\leftrightarrow |e_A\rangle$ transition of atom $A$, while $\Gamma_{B,g}$ and $\Gamma_{B,f}$ correspond to the $|g_B\rangle\leftrightarrow |e_B\rangle$ and $|f_B\rangle\leftrightarrow |e_B\rangle$ transitions of atom $B$, respectively. Under the flat-band approximation, the group velocity and local coupling strengths are frequency-independent within the considered window, rendering $\Gamma_A$, $\Gamma_{B,g}$, and $\Gamma_{B,f}$ constant~\cite{Liao2016Review}.

Besides, we adopt the Markovian approximation, requiring the photon propagation time between adjacent coupling points to be much shorter than the atomic excited-state lifetimes, i.e., $\tau_n = (d_{n+1} - d_n)/v_g\ll 1/\Gamma_{A}, 1/\Gamma_{B,g}, 1/\Gamma_{B,f}$. This time-scale separation eliminates the memory effect of the atoms, allowing the delayed feedback to be reduced to an instantaneous process. Consequently, the frequency-induced variation of the propagation phases is negligible, and $\phi_{2n-1} = k_g (d_{n+1} - d_n)$ and $\phi_{2n} = k_f (d_{n+1} - d_n)$ can be treated as constants~\cite{Kockum2014GiantAtom,Guo2017GiantAtom,Kockum2018DecoherenceFree,Kannan2020GiantAtoms,Feng2021TwoGiantAtoms,CaiJia2021GiantScattering,Sinha2020NonMarkovian}. 

It is worth noting that within the single-excitation subspace introduced previously, the discrete subspace associated with atomic excitations is spanned by the three basis states $\vert 0, e_A, g_B \rangle$, $\vert 0, e_A, f_B \rangle$, and $\vert 0, g_A, e_B \rangle$. The atomic excitation amplitudes corresponding to these states satisfy a three-dimensional linear response equation, which yields the common denominator $\mathcal{D}(\Delta)$ in Eq.~\eqref{eq:D_common} as the cubic determinant of the $3 \times 3$ effective coefficient matrix. Consequently, the equation $\mathcal{D}(\Delta) = 0$ dictates three complex resonance poles that govern the system's scattering dynamics~\cite{Rotter2009NonHermitian}. In the decoupled limit---specifically, when $\phi_1$ or $\phi_5$, and $\phi_2$ or $\phi_6$ are odd multiples of $\pi$---the interatomic coupling terms vanish ($\mathcal{G}_g = \mathcal{G}_f = 0$). This decoupling reduces $\mathcal{D}$ to $\mathcal{E}_g \mathcal{E}_f \mathcal{E}_B$, where the conditions $\mathcal{E}_{g,f,B} = 0$ specify the uncoupled resonances corresponding to the those bare atomic states $\vert 0, e_A, g_B \rangle$, $\vert 0, e_A, f_B \rangle$, and $\vert 0, g_A, e_B \rangle$, respectively. Conversely, for nonzero $\mathcal{G}_{g,f}$, the phase-dependent resonance shifts and the superposition of these poles generate the intricate multi-peak interference observed in the spectrum.

\section{Phase-Selected Frequency Conversion via Local Single-Pole Fano Resonance}
\label{sec:SPA}

Although the exact scattering amplitudes in Sec.~\ref{sec:model} are fully solved, their governing by three complex resonance poles leads to intricate multi-peak interference, obscuring the physical mechanism for efficient frequency conversion. To circumvent this, we construct a local single-pole approximation around the inelastic transmission peak, reducing the spectrum to a transparent local Fano model. 
This model decomposes the inelastic transmission amplitude into a background and a resonant, revealing their interplay is controlled by internal propagation phases. By identifying the background suppression condition and formulating a phase-selection criterion based on the resonance weight, we establish phase-controlled local Fano interference as the fundamental paradigm for high-efficiency conversion. We focus exclusively on the inelastic transmission channel because its interference configuration clearly captures the superposition of direct forward output (background) and atom-mediated reabsorption-reemission (resonance), yielding a highly tunable local Fano lineshape; additionally, under optimal phases, inelastic reflection is strongly suppressed, ensuring unidirectional forward conversion.

For simplicity, we impose the equal-spacing and symmetric-coupling conditions, namely $\phi_1=\phi_3=\phi_5$ and $\phi_2=\phi_4=\phi_6$, and $\Gamma_A=\Gamma_{B,g}=\Gamma_{B,f}\equiv\Gamma$. In this case, the inelastic transmission  amplitude $t_f$ in Eq.~\eqref{eq:scattering_amplitudes} becomes
\begin{widetext}
\begin{equation}
t_f=-\frac{\Gamma e^{i(2\phi_1-3\phi_2)}(1+e^{i\phi_1})(1+e^{i\phi_2})(\Delta-2\Gamma\sin\phi_1)\left[\Gamma e^{3i\phi_2}(1+e^{i\phi_2})^2+i\mathcal{E}_f\right]}
{\mathcal{E}_g\mathcal{E}_f\mathcal{E}_B+\mathcal{G}_g\mathcal{E}_f+\mathcal{G}_f\mathcal{E}_g} .
\label{eq:tf_simplified}
\end{equation}
\end{widetext}

\begin{figure*}[!t]
\centering
\includegraphics[width=0.9\textwidth]{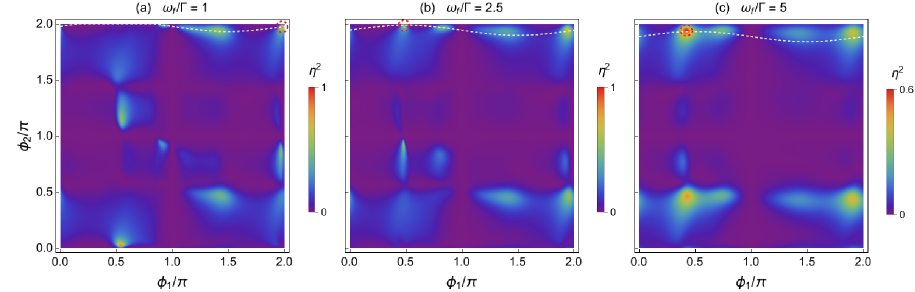}
\caption{Distribution of the single-pole resonance weight $\eta^2$ in the phase plane for varying $\omega_f$. Bright regions indicate optimal phase-selection zones for efficient forward conversion. The white dashed line traces the approximate phase relation derived from the local-background suppression condition near $\phi_2 \simeq 2\pi$.}
\label{fig:PSW-Map}
\end{figure*}

\begin{figure*}[!t]
\centering
\includegraphics[width=0.9\textwidth]{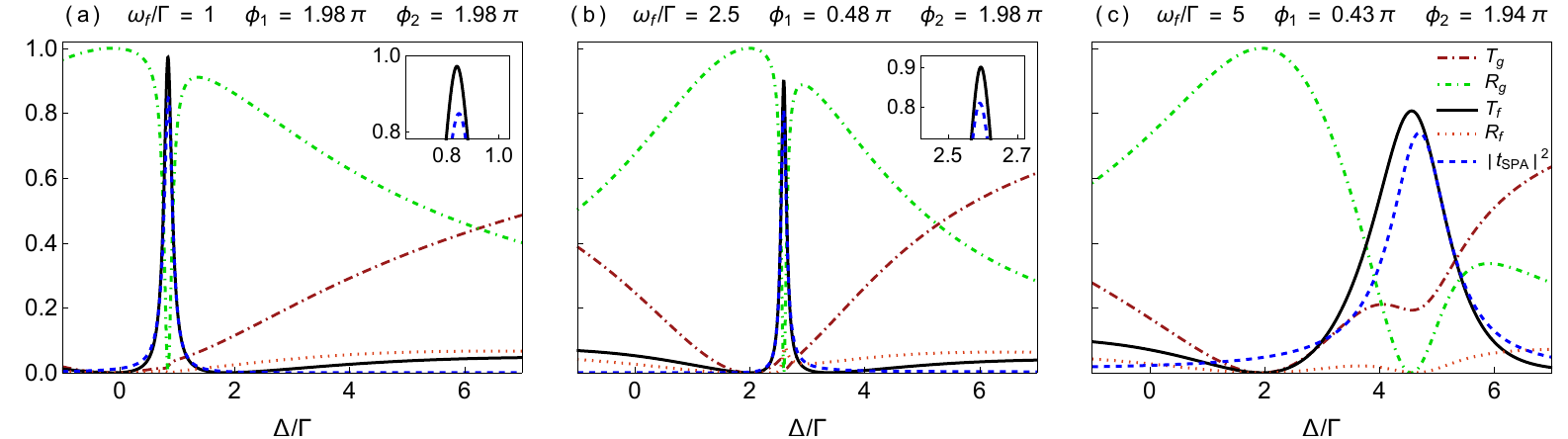}
\caption{Exact four-channel scattering probabilities and the local single-pole approximation $|t_{\mathrm{SPA}}|^2$ for bright-region phases. The approximation accurately captures the target inelastic peak $T_f$ within the target window and confirms the suppression of elastic ($T_g$, $R_g$) and inelastic reflection ($R_f$) channels.}
\label{fig:4CH-Fano}
\end{figure*}

Unlike in a single three-level atom system where the converted photon exits directly~\cite{Tsoi2009LambdaScattering,Witthaut2010ThreeLevelEmitter,Bradford2012FrequencyConversion,Wang2014FrequencyConverter,Du2021GiantLambdaConversion,Zou2022GiantLambdaSQUID}, the present two-giant-atom configuration allows the $f$-channel photon emitted by atom $B$ to either exit directly or be reabsorbed by atom $A$. The coherent superposition of this background continuum (direct output) and the discrete resonant state (reabsorption-induced oscillation by $A$) gives rise to the local Fano lineshape in the inelastic transmission window. The discrete excited state $|0,e_A,f_B\rangle$ formed by this secondary coherent coupling has a bare energy of $\omega_e+\omega_f$. The effective detuning of the total energy $E$ of the single-excitation scattering state relative to this bare energy is
\begin{equation}
    X=E-(\omega_e+\omega_f)=\Delta-\omega_f .
    \label{eq:X_def}
\end{equation}
Accordingly, the inelastic transmission peak near $\Delta\simeq\omega_f$ is identified as the target peak, and the scattering window containing it is defined as the target scattering window. 
For notational simplicity, unless $\Gamma$ is explicitly retained, all frequencies, detunings, linewidths, and loss rates in the following analysis are measured in units of $\Gamma$, i.e., $\Delta/\Gamma \rightarrow \Delta$, $X/\Gamma \rightarrow X$, $\omega_f/\Gamma \rightarrow \omega_f$, $\delta_L/\Gamma \rightarrow \delta_L$, $\gamma_{\text{eff}}/\Gamma \rightarrow \gamma_{\text{eff}}$, and $\kappa/\Gamma \rightarrow \kappa$. The coherent coupling terms are similarly rescaled as $G_{g,f}/\Gamma^2 \rightarrow G_{g,f}$.
By substituting $\Delta=X+\omega_f$ into Eq.~\eqref{eq:tf_simplified} and applying a first-order Taylor expansion to the scattering amplitude’s numerator and denominator around the target peak ($X=0$), the complex three-pole scattering spectrum is reduced to a local single-pole model,
\begin{equation}
t_{\mathrm{SPA}}(X)=\frac{\mathcal{N}^{(0)}+\mathcal{N}^{(1)}X}{\mathcal{D}^{(0)}+\mathcal{D}^{(1)}X}.
\label{eq:t_spa_ND}
\end{equation}
Here, $\mathcal{N}^{(0)}$ and $\mathcal{D}^{(0)}$ are the constant terms of the expansion, while $\mathcal{N}^{(1)}$ and $\mathcal{D}^{(1)}$ are the corresponding linear coefficients:
\begin{equation}
\begin{aligned}
\mathcal{N}^{(0)}&=-e^{i(2\phi_1-3\phi_2)}\mu_1\mu_2(\omega_f-2\sin\phi_1)\left(e^{3i\phi_2}\mu_2^2-2\mu_2\right),\\
\mathcal{N}^{(1)}&=-e^{i(2\phi_1-3\phi_2)}\mu_1\mu_2\bigl[e^{3i\phi_2}\mu_2^2-2\mu_2+i(\omega_f-2\sin\phi_1)\bigr],\\
\mathcal{D}^{(0)}&=(\omega_f+2i\mu_1)(2i\mu_2)(\omega_f+2i\mu_1+2i\mu_2)\\
&\quad+\mathcal{G}_g(2i\mu_2)+\mathcal{G}_f(\omega_f+2i\mu_1),\\
\mathcal{D}^{(1)}&=(2i\mu_2)(\omega_f+2i\mu_1+2i\mu_2)\\
&\quad+(\omega_f+2i\mu_1)(\omega_f+2i\mu_1+2i\mu_2)\\
&\quad+(\omega_f+2i\mu_1)(2i\mu_2)+\mathcal{G}_g+\mathcal{G}_f,
\end{aligned}
\label{eq:spa_coefficients}
\end{equation}
with 
\begin{equation}
\mu_1=1+e^{i\phi_1},\quad\mu_2=1+e^{i\phi_2}.
\label{eq:mu_def}
\end{equation}
Then the effective single-pole position is determined by the zero of the denominator in Eq.~\eqref{eq:t_spa_ND}:
\begin{equation}
X_p=-\frac{\mathcal{D}^{(0)}}{\mathcal{D}^{(1)}}\equiv\delta_{\mathrm L}-i\gamma_{\mathrm{eff}},
\label{eq:xp_def}
\end{equation}
where $\delta_{\mathrm L}$ and $\gamma_{\mathrm{eff}}$ denote the effective Lamb shift and decay rate, respectively. The target scattering window, $W=\{X\vert\vert X-\delta_{\mathrm L}\vert\lesssim\gamma_{\mathrm{eff}}\}$, characterizes the main response range near the target peak.

To isolate the resonant contribution from the direct scattering background and reveal the underlying Fano interference, we recast Eq.~\eqref{eq:t_spa_ND} as
\begin{equation}
t_{\mathrm{SPA}}(X)= t_{\mathrm{bg}}+ \frac{C}{X-X_p},
\label{eq:fano_decomposition}
\end{equation}
with the local background term $t_{\mathrm{bg}}=\mathcal{N}^{(1)}/\mathcal{D}^{(1)}$ and the resonant amplitude $C=[\mathcal{N}^{(0)}\mathcal{D}^{(1)}-\mathcal{N}^{(1)}\mathcal{D}^{(0)}]/(\mathcal{D}^{(1)})^2$. The residual effects of non-target resonances are encapsulated within $t_{\mathrm{bg}}$, $C$, and $X_p$.

Physically, $X$ denotes the effective detuning of the converted photon from the $|0,g_A,f_B\rangle \leftrightarrow |0,e_A,f_B\rangle$ transition. Within the target window $|X-\delta_{\mathrm{L}}|\lesssim\gamma_{\mathrm{eff}}$, the resonant term dominates; outside this window, the secondary coherent coupling is detuned, and $t_{\mathrm{SPA}}(X)$ asymptotically approaches $t_{\mathrm{bg}}$. Because Eq.~\eqref{eq:fano_decomposition} shares the structure of the standard Fano amplitude—a superposition of a background continuum and a discrete resonant state—the inelastic transmission spectrum within the target window exhibits a local Fano lineshape~\cite{Fano1961Configuration,Fan2003FanoCMT,Miroshnichenko2010Fano,Limonov2017Fano,Feng2021TwoGiantAtoms,Yin2022GiantMolecule}.

Defining the auxiliary functions $Q(\phi_1)=\omega_f-2\sin\phi_1$ and $B(\phi_2)=e^{3i\phi_2}\mu_2^2-2\mu_2$, the local background amplitude follows from Eqs.~\eqref{eq:spa_coefficients} and \eqref{eq:fano_decomposition} as
\begin{equation}
\vert t_{\mathrm{bg}}\vert=\frac{\vert\mu_1\mu_2\vert\vert B(\phi_2)+iQ(\phi_1)\vert}{\vert\mathcal{D}^{(1)}\vert},
\label{eq:tbg_abs}
\end{equation}
yielding the local-background suppression condition $B(\phi_2)+iQ(\phi_1)\simeq0$. When suppressed, Eq.~\eqref{eq:fano_decomposition} dictates that $t_{\mathrm{SPA}}$ within the target window is dominated by the single-pole resonant term, which peaks at $|C|^2/\gamma_{\mathrm{eff}}^2$ for $X=\delta_{\mathrm{L}}$. This motivates the single-pole resonance weight:
\begin{equation}
\eta^2=\frac{|C|^2}{\gamma_{\mathrm{eff}}^2}=\frac{|\mu_1\mu_2|^2\,|\mathcal{F}_{\eta}|^2}
{|\mathcal{S}_{\eta}|^2},
\label{eq:eta_compact_ch4}
\end{equation}
with $\mathcal{F}_{\eta}=[B(\phi_2)+iQ(\phi_1)]\mathcal{D}^{(0)}-Q(\phi_1)B(\phi_2)\mathcal{D}^{(1)}$ and $|\mathcal{S}_{\eta}|=\gamma_{\mathrm{eff}}|\mathcal{D}^{(1)}|^2$. Here, $|\mu_1\mu_2|$ is the phase-dictated coherent-coupling envelope, $\mathcal{F}_{\eta}$ is the resonant amplitude factor, and $|\mathcal{S}_{\eta}|\propto\gamma_{\mathrm{eff}}$ characterizes the decay-rate limitation on the resonant response.

Figure~\ref{fig:PSW-Map} maps $\eta^2$ across the phase plane for varying $\omega_f$. The bright regions, indicating a strong single-pole resonant response, serve as optimal phase-selection zones for efficient forward conversion. The prominent bright region used for phase selection lies near $\phi_2 \simeq 2\pi$. Expanding the background suppression condition ($B(\phi_2)+iQ(\phi_1)\simeq0$) around $\phi_2 \simeq 2\pi$ yields the approximate phase relation $\phi_2 \simeq 2\pi - (\omega_f - 2\sin\phi_1)/14$, traced by the white dashed line. Along this curve, $|\mu_2| = |1+e^{i\phi_2}| \simeq 2$ approaches its maximum, and the background is approximately suppressed, providing a necessary condition for the bright region. Conversely, near $\phi_1 \simeq \pi$ or $\phi_2 \simeq \pi$, the coherent-coupling envelope $|\mu_1\mu_2|$ vanishes, collapsing $\eta^2$ and producing the dark regions.

Figure~\ref{fig:4CH-Fano} shows the exact four-channel scattering probabilities ($T_g=|t_g|^2$, $R_g=|r_g|^2$, $T_f=|t_f|^2$, $R_f=|r_f|^2$) together with the local single-pole approximation result $|t_{\mathrm{SPA}}|^2$ for the bright-region phases selected in Fig.~\ref{fig:PSW-Map}. 
Within the target window, $|t_{\mathrm{SPA}}|^2$ closely matches $T_f$ in Figs.~\ref{fig:4CH-Fano}(a) and (b). As $\omega_f$ changes, the target scattering window shifts position; for the larger conversion interval $\omega_f=5\Gamma$ in Fig.~\ref{fig:4CH-Fano}(c), the target peak broadens. While the approximation still captures the peak position and main lineshape, the deviation from exact $T_f$ increases as the broadened linewidth amplifies the relative impact of residual contributions from higher-order terms and non-target resonances. 
Across all cases, the exact spectra confirm that the $\eta^2$-selected phases effectively enhance $T_f$ while suppressing the $T_g$, $R_g$, and $R_f$ channels. Notably, at the phase maximizing $T_f$, destructive interference drives $R_f$ to near zero, confirming that the high-efficiency frequency conversion is predominantly unidirectional in the forward direction.

\begin{figure}[!t]
\centering
\includegraphics[width=0.9\columnwidth]{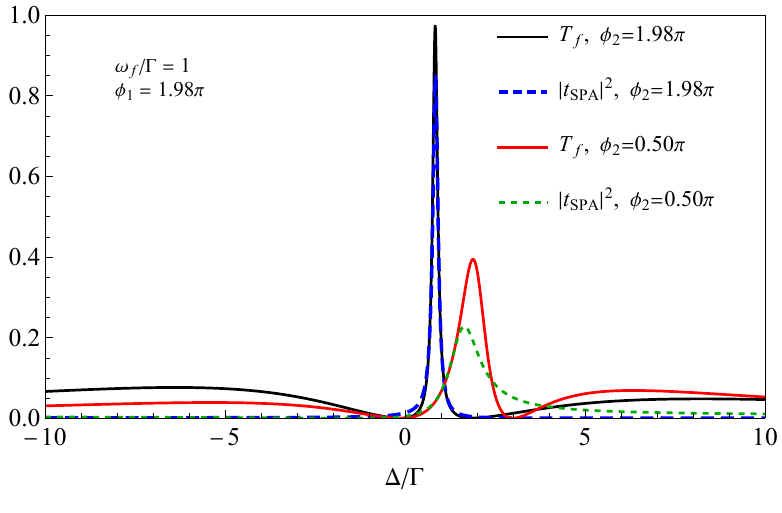}
\caption{Comparison between the exact $T_f$ (solid) and the local single-pole approximation $|t_{\mathrm{SPA}}|^2$ (dashed) for $\omega_f/\Gamma=1$ and $\phi_1=1.98\pi$. Black and blue curves: bright-region phase $\phi_2=1.98\pi$, validating the approximation; red and green curves: reference phase $\phi_2=0.50\pi$, where deviations arise from non-target resonances and background.}
\label{fig:SPA-Validity}
\end{figure}

To further illustrate the phase-dependent validity of the local single-pole approximation, we fix $\omega_f/\Gamma=1$ and $\phi_1=1.98\pi$, and compare the exact inelastic transmission probability $T_f$ with the approximation $|t_{\mathrm{SPA}}|^2$ for two phases: the bright-region phase $\phi_2=1.98\pi$ and a reference phase $\phi_2=0.50\pi$ far from the bright region. As shown in Fig.~\ref{fig:SPA-Validity}, the approximation accurately captures the position and lineshape of the exact peak for the bright phase, indicating that the target window is dominated by the single-pole contribution. In contrast, for the reference phase, significant deviations appear between the exact and approximate spectra, revealing that higher-order terms, background contributions, and non-target resonances become important away from the bright region. Thus, the single-pole approximation is phase-selective: it effectively describes the target peak only near the bright phases identified by $\eta^2$.

\begin{figure}[!t]
\centering
\includegraphics[width=0.9\columnwidth]{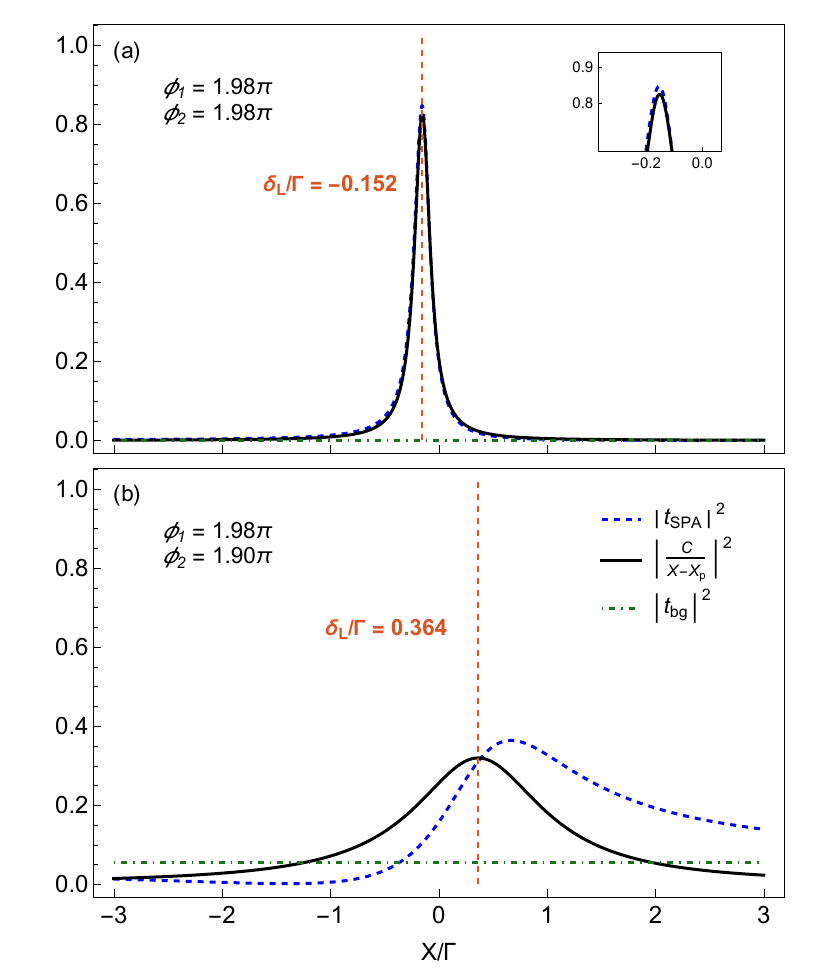}
\caption{Fano decomposition of the local single-pole model near the target peak for $\omega_f/\Gamma=1$. The orange dashed line marks $X = \delta_{\mathrm{L}}$. (a)~$\phi_1=1.98\pi$, $\phi_2=1.98\pi$: suppressed background leaves $|t_{\mathrm{SPA}}|^2$ resonant-term dominated, approaching a Lorentzian lineshape. (b)~$\phi_1=1.98\pi$, $\phi_2=1.90\pi$: non-negligible $|t_{\mathrm{bg}}|^2$ superposes with $|C/(X-X_p)|^2$, yielding a Fano lineshape.}
\label{fig:Fano-Decomp}
\end{figure}

Figure~\ref{fig:Fano-Decomp} decomposes the local single-pole approximation amplitude, Eq.~\eqref{eq:fano_decomposition}, into the local background and single-pole resonant terms for two phase pairs. For the bright-region phases $\phi_1=1.98\pi$, $\phi_2=1.98\pi$ in Fig.~\ref{fig:Fano-Decomp}(a), which is the same as the bright-region phase pair in Fig.~\ref{fig:4CH-Fano}(a), the suppressed background leaves the expression dominated by the resonant term. Thus, $|t_{\mathrm{SPA}}|^2$ nearly coincides with $|C/(X-X_p)|^2$, approaching a Lorentzian lineshape with its window center approximately at $X=\delta_{\mathrm{L}}$ and linewidth characterized by $\gamma_{\mathrm{eff}}$. For $\phi_1=1.98\pi$, $\phi_2=1.90\pi$ in Fig.~\ref{fig:Fano-Decomp}(b), the non-negligible $|t_{\mathrm{bg}}|^2$ coherently superposes with $|C/(X-X_p)|^2$, making $|t_{\mathrm{SPA}}|^2$ deviate markedly from the resonant term and yielding a local Fano lineshape; consequently, $\delta_{\mathrm{L}}$ characterizes only the resonant-term center rather than the total spectrum.

In summary, Sec.~\ref{sec:SPA} establishes that high-efficiency frequency conversion is dictated by phase-controlled local Fano interference. The internal propagation phases $\phi_1$ and $\phi_2$ exclusively govern the background-resonance interplay: bright-region phases suppress the background, leaving the resonance dominant and yielding a Lorentzian with unidirectional forward conversion. Crucially, this internal phase control utilizes giant-atom multipath interference to deterministically isolate the resonance, establishing a general paradigm for engineering resonant light-matter interactions in structured quantum systems.

\FloatBarrier
\section{Performance Advantage and Robustness }
\label{sec:results}

\begin{figure}[!t]
    \centering
    \includegraphics[width=0.9\columnwidth]{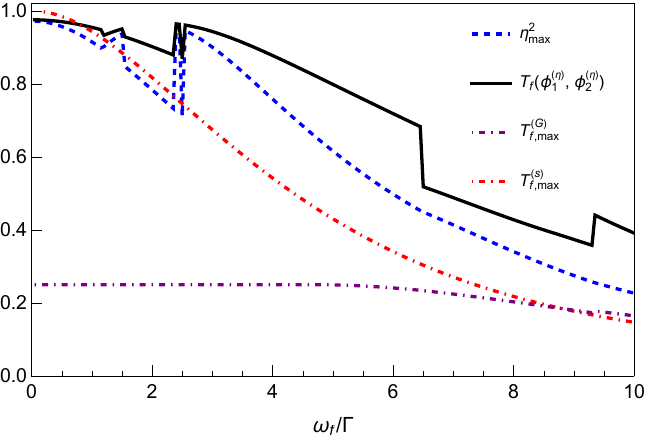}
\caption{Advantage of the two-giant-atom scheme over the single $\Lambda$-type giant-atom and small-atom models. The similar trends of $\eta_{\max}^2$ and $T_f(\phi_1^{(\eta)}, \phi_2^{(\eta)})$ validate $\eta^2$ as an effective phase-selection criterion.} 
    \label{fig:Scaling-Curve}
\end{figure}

Having established the phase-controlled local Fano interference mechanism in Sec.~\ref{sec:SPA}, we now comprehensively evaluate the system's performance. We first contrast the present system with the single $\Lambda$-type giant-atom and small-atom models to reveal how secondary coherent coupling and internal phase control reconstruct the pole equation and enhance the inelastic channel. Subsequently, we assess the system's robustness against excited-state loss, demonstrating the persistence of this enhancement under realistic non-Hermitian conditions. This combined analysis confirms that the interplay of secondary coherent coupling and internal phase control enables substantially enhanced and robust frequency conversion over a broad range.

To characterize the influence of the frequency-conversion interval $\omega_{f}$ on the phase distribution of $\eta^2$, we define
\begin{equation}
\begin{aligned}
& (\phi_1^{(\eta)}, \phi_2^{(\eta)}) = \underset{\phi_1, \phi_2 \in [0, 2\pi)}{\arg\max} \, \eta^2(\phi_1, \phi_2; \omega_f), \\
& \eta_{\max}^2 = \eta^2(\phi_1^{(\eta)}, \phi_2^{(\eta)}; \omega_f), \\
& T_f(\phi_1^{(\eta)}, \phi_2^{(\eta)}) = \underset{X \in W}{\max} \, T_f(X; \phi_1^{(\eta)}, \phi_2^{(\eta)}, \omega_f).
\end{aligned}
\end{equation}

Equipped with this phase-selection criterion, we now assess the advantage of the two-giant-atom scheme through systematic model comparison.
To evaluate the role of secondary coherent coupling, we set $\Gamma_A=0$ to reduce the system to a single $\Lambda$-type giant-atom model~\cite{Du2021GiantLambdaConversion,Zou2022GiantLambdaSQUID}, where the two-level giant atom $A$ decouples and the $\Lambda$-type giant atom $B$ solely provides the frequency-conversion channel. Defining the propagation phases as $\phi_1^{(\mathrm G)}=k_g(d_4-d_3)$ and $\phi_2^{(\mathrm G)}=k_f(d_4-d_3)$, the inelastic transmission probability under symmetric coupling follows from Eqs.~\eqref{eq:D_common} and \eqref{eq:numerators} as
\begin{equation}
T_f^{(\mathrm{G})}(X) = \frac{\Gamma^2 |\mu_1^{(\mathrm{G})} \mu_2^{(\mathrm{G})}|^2}{|\mathcal{D}^{(\mathrm{G})}(X)|^2},
\end{equation}
where
\begin{equation}
\begin{aligned}
& \mathcal{D}^{(\mathrm{G})}(X) = \omega_f + X + 2i\Gamma\mu_1^{(\mathrm{G})} + 2i\Gamma\mu_2^{(\mathrm{G})}, \\
& \mu_1^{(\mathrm{G})} = 1 + e^{i\phi_1^{(\mathrm{G})}}, \quad \mu_2^{(\mathrm{G})} = 1 + e^{i\phi_2^{(\mathrm{G})}}.
\end{aligned}
\end{equation}
The maximum of $T_f^{(\mathrm{G})}$ within the target scattering window is designated as $T_{f,\max}^{(\mathrm{G})}$. In the near-degenerate limit $\omega_f \to 0$, at the effective resonance and under symmetric coupling, a single $\Lambda$-type giant atom partitions the scattering probability equally among the four channels ($T_g=R_g=T_f = R_f = 0.25$) ~\cite{Du2021GiantLambdaConversion}; whereas as $\omega_f$ increases, $T_{f,\max}^{(\mathrm{G})}$ remains slightly below $0.25$ due to the propagation-phase mismatch $\Delta\phi = \omega_f(d_4-d_3)/v_g$ between the $g$- and $f$-channels. Unlike this decoupled limit, the present system retains secondary coherent-coupling terms proportional to $\Gamma_A$, which fundamentally reconstruct the inelastic transmission amplitude. Specifically, the $\Gamma_A$-dependent terms in $\mathcal{G}_{g}$ and $\mathcal{G}_{f}$ shift the pole positions and effective linewidths beyond the sole determination of $\mathcal{D}^{(\mathrm{G})}(X)$, while the corresponding numerator terms in Eq.~\eqref{eq:numerators} render the target peak strength explicitly dependent on the propagation phases. Crucially, as shown in Fig.~\ref{fig:4CH-Fano}, this secondary coupling introduces an additional phase-tunable interference pathway that breaks the original symmetric decay, enabling constructive interference exclusively in the forward inelastic channel while suppressing the others. As confirmed in Fig.~\ref{fig:Scaling-Curve}, $T_f(\phi_1^{(\eta)},\phi_2^{(\eta)})$ significantly exceeds $T_{f,\max}^{(\mathrm{G})}$ over a broad $\omega_f$ range, demonstrating that phase-controlled secondary coherent coupling effectively enhances the inelastic transmission channel.

To evaluate the phase-control effect introduced by the giant-atom structure, we introduce a small-atom model by contracting each pair of coupling points into a single point ($d_A$ and $d_B$)~\cite{Kockum2014GiantAtom,Kockum2021GiantAtomsReview,Du2021GiantLambdaConversion}, with single-point decay rates $\Gamma_A^{(s)} = \Gamma_{B,g}^{(s)} = \Gamma_{B,f}^{(s)} = 2\Gamma$ and propagation phases $\phi_1^{(s)}=k_g(d_B-d_A)$ and $\phi_2^{(s)}=k_f(d_B-d_A)$. Using the real-space approach, the inelastic transmission probability is obtained as
\begin{equation}
T_f^{(s)}(X) = \frac{4\Gamma^2 (\omega_f + X)^2 |X + 2i\Gamma - 2i\Gamma e^{2i\phi_2^{(s)}}|^2}{|\mathcal{D}^{(s)}(X)|^2},
\label{eq:Tf_small}
\end{equation}
where
\begin{equation}
\begin{aligned}
\mathcal{D}^{(s)}(X) = & \, (\omega_f + X + 2i\Gamma)(X + 2i\Gamma)(\omega_f + X + 4i\Gamma) \\
& + 4\Gamma^2 e^{2i\phi_1^{(s)}} (X + 2i\Gamma) \\
& + 4\Gamma^2 e^{2i\phi_2^{(s)}} (\omega_f + X + 2i\Gamma).
\end{aligned}
\label{eq:D_small}
\end{equation}
The maximum of $T_f^{(s)}$ within the target scattering window is denoted by $T_{f,\max}^{(s)}$. The advantage of the present system over the small-atom model lies in how phases enter the scattering amplitudes. In the small-atom model, the phase factors $e^{2i\phi_{1,2}^{(s)}}$ control the response solely through the external propagation phases between $d_A$ and $d_B$. In contrast, for the giant-atom model, the internal coherent factors $\mu_1$ and $\mu_2$ simultaneously permeate the coherent-coupling envelope, the common denominator $\mathcal{D}$, and the local single-pole parameters. This internal phase control enables the giant-atom structure to maintain a high inelastic transmission probability over a broader $\omega_f$ range. 
The apparent superiority of the small-atom model over the giant-atom model near $\omega_f \to 0$ in Fig.~\ref{fig:Scaling-Curve} arises solely from the equal-spacing constraint adopted for analytical tractability. Without this constraint, the giant-atom model reduces to a more general framework with the small-atom model as a special case, inherently capable of surpassing the small-atom limit across the entire frequency domain.

In the large-detuning limit ($\omega_f \gg \Gamma, X$), substituting $\Delta = \omega_f + X$ into Eq.~\eqref{eq:tf_simplified} and extracting the leading-order terms yields the asymptotic scaling of the inelastic transmission probability for the giant-atom model:
\begin{equation}
T_f(X) \sim \frac{\Gamma^{2}|\mu_1\mu_2|^2}{\omega_f^2} \frac{|\Gamma B(\phi_2) + iX|^2}{|X + 2i\Gamma\mu_2|^2}.
\label{eq:tf_large_wf}
\end{equation}
Applying the same procedure to Eq.~\eqref{eq:Tf_small}, the small-atom model exhibits an analogous scaling:
\begin{equation}
T_f^{(s)}(X) \sim \frac{4\Gamma^{2}}{\omega_f^2} \frac{|X + 2i\Gamma(1 - e^{2i\phi_2^{(s)}})|^2}{|X + 2i\Gamma|^2}.
\label{eq:tf_small_large_wf}
\end{equation}
Equations~\eqref{eq:tf_large_wf} and \eqref{eq:tf_small_large_wf} reveal that both small-atom and giant-atom models are subject to a universal $1/\omega_f^2$ decay as the conversion interval increases, dictated by the dominant $\omega_f$ terms in the denominators. While phase control cannot alter this global scaling, it critically governs the bounded phase factors and the local Fano lineshape, thereby maximizing the peak height within the target scattering window. Figure~\ref{fig:Scaling-Curve} compares $\eta_{\max}^2$, $T_f(\phi_1^{(\eta)}, \phi_2^{(\eta)})$, $T_{f,\max}^{(s)}$, and $T_{f,\max}^{(\mathrm{G})}$ as functions of $\omega_f$. The similar overall trends of $\eta_{\max}^2$ and $T_f(\phi_1^{(\eta)}, \phi_2^{(\eta)})$ validate $\eta^2$ as an effective phase-selection criterion, with minor numerical deviations arising from residual background and non-target resonant contributions. The local jumps originate from the transfer of $(\phi_1^{(\eta)}, \phi_2^{(\eta)})$ between different bright regions as $\omega_f$ varies.

\begin{table}[!t]
\centering
\caption{
Values of $\widetilde T_{f,\max}(\kappa)$ for different parameters.
}
\label{tab:loss_robustness}
\begin{tabular}{c c c c c}
\hline\hline
$\omega_f/\Gamma$ 
& $\kappa/\Gamma=0$ 
& $\kappa/\Gamma=0.02$ 
& $\kappa/\Gamma=0.05$ 
& $\kappa/\Gamma=0.10$ \\
\hline
$1$   & $0.973$ & $0.802$ & $0.690$ & $0.642$ \\
$2.5$ & $0.974$ & $0.853$ & $0.801$ & $0.731$ \\
$5$   & $0.822$ & $0.795$ & $0.759$ & $0.705$ \\
\hline\hline
\end{tabular}
\end{table}

Finally, to test the robustness of the system against loss, we introduce a non-Hermitian correction $\omega_e \rightarrow \omega_e - i\kappa/2$ to the excited-state energies~\cite{Rotter2009NonHermitian}. This modifies the complex detunings in Eq.~\eqref{eq:auxiliary_definitions} as $\mathcal{E}_{g,f,B} \rightarrow \mathcal{E}_{g,f,B} + i\kappa/2$, and correspondingly replaces $\Delta - 2\Gamma\sin\phi_1$ in the numerator of Eq.~\eqref{eq:tf_simplified} with $\Delta + i\kappa/2 - 2\Gamma\sin\phi_1$. Within the target scattering window, the loss-affected maximum inelastic transmission probability is defined as
\begin{equation}
    \widetilde T_{f,\max}(\kappa) = \max_{\phi_1,\phi_2\in[0,2\pi),\,X\in W} \left| \widetilde t_f \left( X;\phi_1,\phi_2,\omega_f,\kappa \right) \right|^2 .
    \label{eq:Tf_loss_max}
\end{equation}
As shown in Table~\ref{tab:loss_robustness}, although loss inevitably degrades the transmission, the system exhibits considerable robustness: for $\kappa/\Gamma \leq 0.10$, $\widetilde T_{f,\max}(\kappa)$ remains above $0.64$ across the three conversion intervals, with values of $0.642$, $0.731$, and $0.705$ for $\omega_f/\Gamma = 1, 2.5, 5$ at $\kappa/\Gamma = 0.10$, respectively. Fundamentally, loss introduces an irreversible probability current outflow into unguided modes; while phase control can partially mitigate its impact on the target channel, it cannot restore the unitary scattering limit of the lossless system.

\section{Summary and Discussion}
\label{sec:conclusion}

We have investigated high-efficiency single-photon frequency conversion in a waveguide-QED structure comprising a two-level giant atom and a $\Lambda$-type three-level giant atom. The $\Lambda$-type atom provides the inelastic channel, while the two-level atom induces secondary coherent coupling, creating multipath interference for the converted photon. The inelastic transmission spectrum, governed by three complex resonance poles, exhibits an intricate multi-peak interference pattern. By constructing a local single-pole approximation around the target peak, we reduced this complex spectrum to a local Fano lineshape, decomposing it into a coherent superposition of a local background term and a single-pole resonant term. Crucially, the interplay between these two terms is exclusively controlled by the internal propagation phases $\phi_{1,2}$: bright-region phases suppress the background, leaving the resonance dominant and yielding a Lorentzian reduction with unidirectional forward conversion.

Based on this decomposition, we formulated the single-pole resonance weight $\eta^2 = |C|^2/\gamma_{\mathrm{eff}}^2$ as a phase-selection criterion for highly efficient conversion. Exact scattering calculations confirm that the $\eta^2$-selected phases significantly enhance the forward inelastic transmission while suppressing all other channels. Comparative analysis reveals the fundamental advantage of the two-giant-atom scheme: unlike the single $\Lambda$-type giant-atom model, secondary coherent coupling reconstructs the pole equation to enable cooperative tuning of the pole position and residue; unlike the small-atom model, internal phase control compensates for the universal $1/\omega_f^2$ asymptotic decay, maintaining high transmission over a broader frequency-conversion range. Furthermore, the system exhibits considerable robustness against excited-state loss, preserving high target transmission for $\kappa/\Gamma \leq 0.10$.

For experimental implementation, the present model can be constructed on a superconducting artificial-atom waveguide-QED platform~\cite{Gu2017MicrowavePhotonics}. The two-level giant atom $A$ can be constructed from two levels of a superconducting artificial atom, and the $\Lambda$-type giant atom $B$ from a three-level subspace of a multilevel artificial atom. Related level-structure engineering, $\Lambda$-type structures in superconducting artificial atoms, and $\Lambda$-type giant-atom schemes have been investigated in superconducting circuits~\cite{Kelly2010CPT,Novikov2016Raman,Vadiraj2021LevelStructure,Zou2022GiantLambdaSQUID}. The spatially separated coupling points can be connected to the same microwave transmission line via multiple coupling ports, with propagation phases set by the distances between coupling points, the operating frequency, and the effective electrical length of the waveguide~\cite{Kockum2014GiantAtom,Kannan2020GiantAtoms}.

The present analysis relies on the Markovian approximation and adopts the equal-spacing and symmetric-coupling conditions. Future work may explore the effects of unequal-spacing coupling, asymmetric coupling, and non-Markovian feedback on the system's performance~\cite{Guo2017GiantAtom,Andersson2019Nonexponential,Sinha2020NonMarkovian,Guo2020Oscillating}. Moreover, the local single-pole approximation and the Fano lineshape decomposition established here provide a general analytical paradigm that can be extended to other multi-pole waveguide-QED scattering systems for analyzing resonant enhancement, background suppression, and channel selection within a target scattering window.

\begin{acknowledgments}
Y. J. Song was supported by the National Natural Science Foundation of China (Grant No. 12205088), the Natural Science Foundation of Hunan Province (Grant Nos. 2026JJ50350, 2025JJ50005), and the Open Project of Key Laboratory of Low Dimensional Quantum Structures and Quantum Control of Ministry of Education of Hunan Normal University (Grant No. QSQC2602). Y. Liu was supported by the Scientific Research Fund of Hunan Provincial Education Department of China (Grant No. 24C0353) and the Open Project of Key Laboratory of Opto-electronic Control and Detection Technology of University of Hunan Province (Grant No. 2024HSKFJJ012).
\end{acknowledgments}

\bibliographystyle{apsrev4-2}
\bibliography{refs}
\end{document}